\documentclass[aip,rsi,amsmath,amssymb,reprint]{revtex4-1}
\usepackage{graphicx}
\usepackage{color}
\usepackage{textcomp}
\usepackage[normalem]{ulem}  
\usepackage{lineno}

\begin{document}

\title{The infrared spectrum of protonated buckminsterfullerene, C$_{60}$H$^{+}$}

\author{Julianna Palot\'as}
\affiliation{Radboud University, Institute for Molecules and Materials, FELIX Laboratory, Toernooiveld 7, 6525ED Nijmegen, The Netherlands}

\author{Jonathan Martens}
\affiliation{Radboud University, Institute for Molecules and Materials, FELIX Laboratory, Toernooiveld 7,
6525ED Nijmegen, The Netherlands}

\author{Giel Berden}
\affiliation{Radboud University, Institute for Molecules and Materials, FELIX Laboratory, Toernooiveld 7,
6525ED Nijmegen, The Netherlands}

\author{Jos Oomens}
\email[Corresponding author: ]{j.oomens@science.ru.nl}
\affiliation{Radboud University, Institute for Molecules and Materials, FELIX Laboratory, Toernooiveld 7,
6525ED Nijmegen, The Netherlands}
\affiliation{van 't Hoff Institute for Molecular Sciences, University of Amsterdam, Science Park 908, 1098XH Amsterdam, The Netherlands}

\begin{abstract}
{\bf Although fullerenes have long been hypothesized to occur in interstellar environments, their actual unambiguous  spectroscopic identification is of more recent date.\cite{cami,sellgren2010,c60GarciaH2010,campbell2015} C$_{60}$, C$_{70}$ and C$_{60}^+$ now constitute the largest molecular species individually identified in the interstellar medium (ISM). Fullerenes have significant proton affinities and it was suggested that C$_{60}$H$^+$ is likely the most abundant interstellar analogue of C$_{60}$.\cite{kroto1992} We present here the first laboratory infrared (IR) spectrum of gaseous C$_{60}$H$^+$. Symmetry breaking relative to C$_{60}$ produces an IR spectrum that is much richer than that of C$_{60}$. The experimental spectrum is used to benchmark theoretical spectra indicating that the B3LYP density functional with the 6-311+G(d,p) basis set accurately reproduces the spectrum. Comparison with IR emission spectra from two planetary nebulae, SMP LMC56 and SMC16, that have been associated with high C$_{60}$ abundances, indicate that C$_{60}$H$^+$ is a plausible contributor to their IR emission.
}     
\end{abstract}

\maketitle

Buckminsterfullerene C$_{60}$ is undoubtedly one of the most iconic molecules of our time. Since its discovery in 1985,\cite{kroto1985} its physico-chemical properties have been extensively characterized, including its ion chemistry and spectroscopic properties. IR spectra have been reported in condensed and gas phases,\cite{c60rempi,fularaC60,schettino2001,Changala2019} and spectra for ionized forms are available as well.\cite{fularaC60,campbell2015,kupser2008,roithova2018} 

The high cosmic abundance of carbon combined with the high stability of fullerenes\cite{fowler1996} initiated a quest for their detection in inter- and circumstellar environments.\cite{somerville1989,snow1989,Ehrenfreund1994,herbig2000} This search culminated in the identifications of neutral C$_{60}$ and C$_{70}$ in a young planetary nebula (Tc1) \cite{cami} based on diagnostic IR features. Accurate gas-phase laboratory spectra in the near-IR range led to the identification of C$_{60}^+$ as carrier of two of the diffuse interstellar bands near 9600 \AA.\cite{campbell2015} 

The question of whether or not fullerenes can form in H-rich regions of the interstellar medium (ISM) has been under debate.\cite{cami,c60GarciaH2010} Hydrogenation produces stable fullerene derivatives and partially hydrogenated fullerenes (fulleranes) have been suggested to occur in circumstellar envelopes and in the ISM \cite{c60GarciaH2010,fullereneic418,zhang2017}. On the other hand, hydrogenation and the concomitant change in orbital hybridization from $sp^2$ to $sp^3$ reduces the stability of the fullerene cage, which under the conditions of the ISM would likely lead to dehydrogenation and restoration of the original fullerene\cite{iglesiasG2012} or to breakdown of the carbon cage.\cite{kroto1992} However, in this latter paper, Kroto also noted that protonation does not compromise cage stability and hypothesized that "protonated C$_{60}$ is likely to be the most abundant fullerene analogue," analogous to high abundances of protonated carbon monoxide, HCO$^+$. 

\begin{figure}[htp]
\begin{center}
\includegraphics[width=0.48\textwidth]{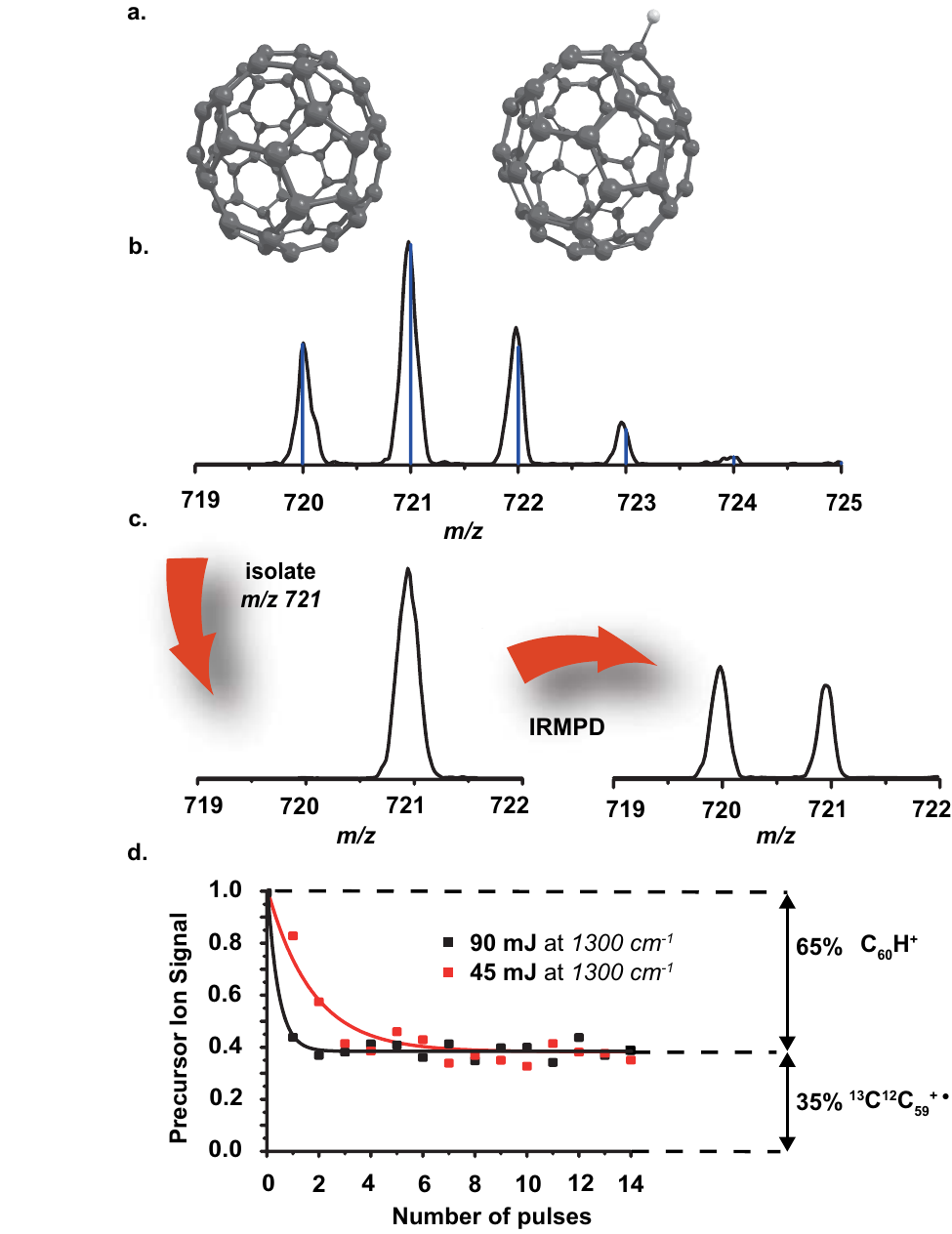}
\caption{\textbf{Structure of C$_{60}$H$^+$ and experimental aspects of recording its IRMPD spectrum. a) Protonation of C$_{60}$ occurs on one of the sixty equivalent C-atoms. b) Mass spectrum as generated by the APCI source without mass isolation. The predicted isotope pattern for a 35\%/65\% mixture of C$_{60}^{+\bullet}$ and C$_{60}$H$^+$ (as represented by the sticks) closely reproduces the experimental pattern. c) Mass spectrum after isolation of the {\textit m/z} 721 peak. With the IR laser frequency on-resonance with an IR absorption of C$_{60}$H$^{+}$ at 1300 cm$^{-1}$, fragmentation by loss of an H-atom occurs leading to an increase of ions in the {\textit m/z} 720 mass channel. d) Increasing the number of IR pulses leads to depletion of the C$_{60}$H$^+$ ion population. Fragmentation of $^{13}$C$^{12}$C$_{59}^{+\cdot}$ is not observed under our experimental conditions, so that the {\textit m/z} 721 peak intensity levels off to a constant value. Note that increasing the  radiation power, saturation is reached sooner, but the limiting value is the same, namely the fraction of $^{13}$C$^{12}$C$_{59}^{+\bullet}$ ions.}}
\label{fig:Fig2}
\end{center}
\end{figure}

Ion chemistry studies\cite{bohme2009} have determined the proton affinity (PA) of C$_{60}$ at 860\,kJ/mol. This relatively high value, just above the PA of ammonia, makes C$_{60}$H$^+$ (Figure \ref{fig:Fig2}a) one of the most relevant stable fullerene derivatives and underpins Kroto's statement above.
However, to our knowledge, no IR spectra have been reported for protonated fullerenes. Electronic spectra have been suggested to suffer from broad, unresolved features due to tunneling of H$^+$ between the identical C-atom sites.\cite{kroto1992} IR spectra have been reported for deposited fullerene films exposed to atomic hydrogen\cite{stoldt2001} and for some stable C$_{60}$H$_{\text{n}}$ hydrogenated fullerenes.\cite{iglesiasG2012} 

Here we present the first experimental IR spectrum of protonated C$_{60}$, recorded in the gas phase via infrared multiple-photon dissociation (IRMPD) using the FELIX free-electron laser and an ion trap mass spectrometer.\cite{oomens2003}
This IR spectrum is compared with astronomical spectra of objects that were associated with high C$_{60}$ abundances as well as with theoretical IR spectra to evaluate the performance of different computational approaches. Contrasting the spectrum with that of C$_{60}$ yields a textbook example of the effects of symmetry breaking on vibrational selection rules.

\begin{figure*}[htp]
\begin{center}
\includegraphics[width=0.8\textwidth]{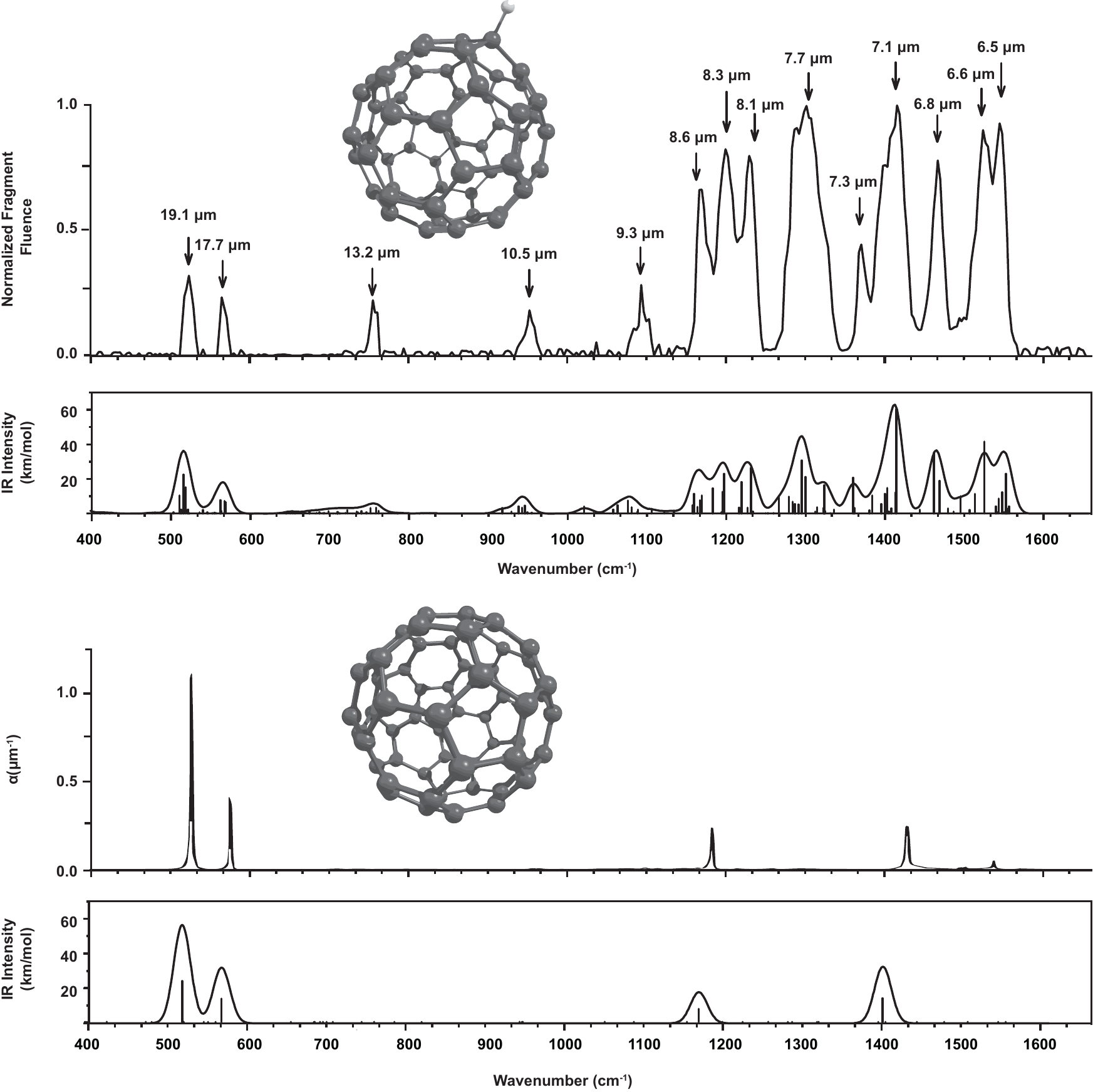}
\caption{\textbf{Experimental IRMPD spectrum of C$_{60}$H$^{+}$ (top) compared with a Fourier transform infrared absorption spectrum of a thin film of neutral C$_{60}$ (bottom) taken from Ref.\ \onlinecite{holleman}. The C$_{60}$ spectrum shows the four IR allowed modes as well as two very weak bands around 1500 cm$^{-1}$, which have been attributed to combination modes. The effects of symmetry breaking on the IR selection rules in C$_{60}$H$^{+}$ are striking. Band positions in $\mu$m are indicated on the IRMPD spectrum. The stick spectra below the experimental data show the theoretical calculations for both molecules at the B3LYP/6-311+G(d,p) level of theory. }}
\label{fig:Fig3}
\end{center}
\end{figure*} 
 
C$_{60}$H$^{+}$ generated by atmospheric-pressure chemical ionization (APCI) produces the mass spectrum in Figure \ref{fig:Fig2}b, showing the radical cation of C$_{60}$ at {\textit m/z} 720, clearly resolved from the higher-intensity mass peak at {\textit m/z} 721. At a natural carbon-13 abundance of 1.1\%, monoisotopic $^{12}$C$_{60}$ occurs at only 51\%, while the $^{13}$C$^{12}$C$_{59}$ isotopomer has an abundance of 34\%. Hence, the {\textit m/z} 721 base peak is a superposition of two ions: the protonated ion C$_{60}$H$^{+}$ and the $^{13}$C$^{12}$C$_{59}^{+\bullet}$ radical cation; the small mass difference of 0.0045 u cannot be resolved in our mass spectrometer. From the observed isotope pattern, the relative populations of C$_{60}$H$^{+}$ and C$_{60}^{+\bullet}$ are derived as 65\% and 35\%, respectively. 

Figure \ref{fig:Fig2}c shows the mass spectrum after isolation of the {\textit m/z} 721 ion in the trap (see Methods). Tuning the laser frequency to a vibrational resonance of C$_{60}$H$^{+}$ leads to the absorption of multiple IR photons inducing dissociation forming the {\textit m/z} 720 ion. Note that only C$_{60}$H$^{+}$ can undergo fragmentation into this channel -- by loss of an H-atom -- while $^{13}$C$^{12}$C$_{59}^{+\bullet}$ cannot. Moreover, due to its high stability, C$_{60}^{+\bullet}$ does not undergo dissociation under our experimental conditions, as was tested by isolating the {\textit m/z} 720 ion. 

Increasing the number of laser pulses, the C$_{60}$H$^{+}$ population undergoes dissociation until it is depleted. The remaining intensity at {\textit m/z} 721 is then exclusively due to $^{13}$C$^{12}$C$_{59}^{+\bullet}$. Figure \ref{fig:Fig2}d confirms its approximately 35\% contribution to the ion population. By monitoring the fragmentation of {\textit m/z} 721 into channel {\textit m/z} 720 as function of wavelength, we selectively measure the IRMPD spectrum of C$_{60}$H$^{+}$, which is not contaminated by contributions from $^{13}$C$^{12}$C$_{59}^{+\bullet}$. 

Buckminsterfullerene C$_{60}$ belongs to the icosahedral point group (I$_\textrm{h}$),\cite{Bethune1990,Changala2019} which leads to IR selection rules that leave the large majority of the 174 normal modes inactive. There are only four triply-degenerate modes belonging to the T$_{1\text{u}}$ irreducible representation that are infrared active. 
As a consequence, C$_{60}$ has an extremely sparse IR spectrum, as reproduced in Figure \ref{fig:Fig3}.\cite{holleman} 

\begin{figure}[htp]
\begin{center}
\includegraphics[width=0.45\textwidth]{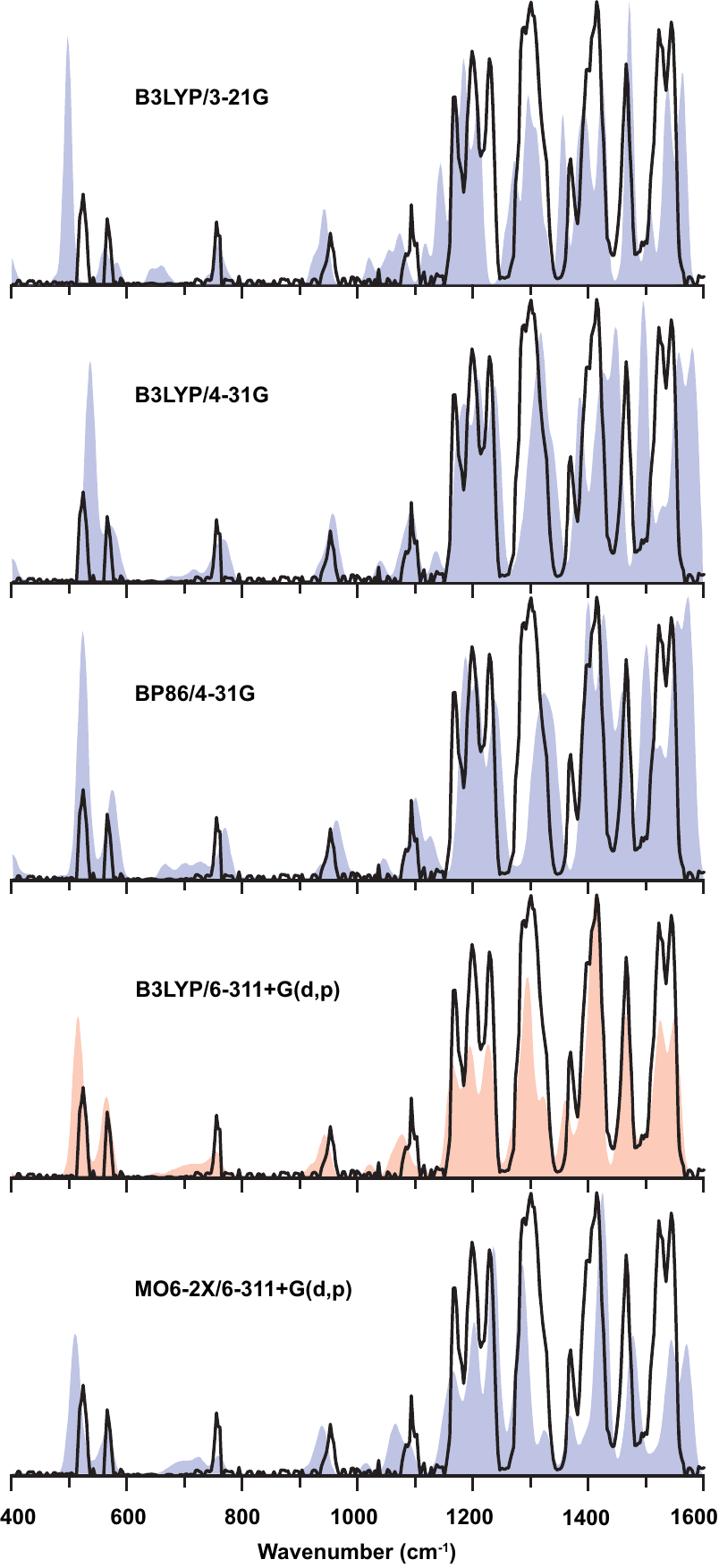}
\caption{\textbf{The experimental IR spectrum of C$_{60}$H$^{+}$ compared with DFT computed spectra using different basis sets and functionals. The best match, marked in red, is obtained using the B3LYP functional with the 6-311+G(d,p) basis set. The RMS deviation between band centers in the experimental spectrum and in the scaled and convoluted theoretical spectra amounts to 1.8 cm$^{-1}$ for this level of theory and is much smaller than for the other computational approaches (see Supplementary Table 1). }}
\label{fig:Fig4}
\end{center}
\end{figure}

Protonation of aromatic systems occurs through {$\sigma$}-bonding to one of the peripheral C-atoms.\cite{Dopfer}  Protonation of C$_{60}$ is therefore expected to occur on one of the 60 C-atoms, and not centrally over one of the 5- or 6-membered rings. Then, since all C-atoms are symmetrically identical, there exists only one isomer of C$_{60}$H$^+$. The C-atoms do not coincide with one of the five-fold rotation axes of C$_{60}$, so that the proton in C$_{60}$H$^+$ is not on a C$_{5}$-axis and thus breaks the five-fold rotation symmetry. In fact, protonation removes all symmetry of the original C$_{60}$ molecule, except for one mirror plane that includes the proton and one of the principal axes of C$_{60}$. Protonation reduces the symmetry to C$_{\textrm{s}}$, in which vibrational modes belong either to A$^{\prime}$ or A$^{\prime \prime}$ irreps, both giving allowed IR transitions. In sharp contrast to C$_{60}$, one thus expects a rich vibrational spectrum for C$_{60}$H$^+$. 

The top panel of Figure \ref{fig:Fig3} presents the IRMPD spectrum of C$_{60}$H$^{+}$ compared with that of C$_{60}$ below it.  The C$_{60}$H$^{+}$ spectrum indeed features a significantly larger number of vibrational bands due to symmetry lowering to C$_{\textrm{s}}$. 

Figure \ref{fig:Fig4} compares the experimental spectrum with computed harmonic spectra at various levels of density functional theory (DFT), as further detailed in the Methods Section.  Figure \ref{fig:Fig4} and Supplementary Table 1 clearly testify that the B3LYP/6-311+G(d,p) method outcompetes all others in the prediction of the IR spectrum. The total integrated intensity of all fundamental vibrational modes at this level of theory is 790 km/mol, versus 215 km/mol for C$_{60}$.

Our statement that there exists only a single isomer of C$_{60}$H$^+$ ignored the -- perhaps naive but intriguing -- possibility of endohedral protonation, {\textit i.e.}\ inside the cage. The optimized geometry is 193\,kJ/mol higher in energy than the exohedrally protonated structure. Moreover, the predicted spectrum is in poorer agreement with the experimental spectrum (see Supplementary Figure 1) and will not be further considered.  

Overall, the C$_{60}$H$^{+}$ spectrum features strong bands in the 1150 -- 1570 cm$^{-1}$ (6.4 -- 8.7 \textmu m) range and weaker ones in the long-wavelength range. This general spectral shape resembles that of ionized PAHs,\cite{allamandola1999} but a detailed inspection reveals significant differences (see Supplementary Figure 2). For C$_{60}$H$^{+}$, the strong bands have mainly CC stretching character, with some of the vibrations near 1165 cm$^{-1}$ having additional CH bending character. For PAHs, the CC-stretch bands extend to shorter wavelengths (6.2 \textmu m).\cite{peeters2002} The C$_{60}$H$^{+}$ bands in the 7.7 -- 8.6 \textmu m range overlap with strong PAH cation bands, but  bands to the blue (6.5 -- 7.1 \textmu m) fall in a relatively silent region of the PAH spectrum.

Towards longer wavelengths, main bands in the PAH spectrum are due to CH out-of-plane vibrations at 11.3, 12.0 and 12.7 \textmu m. The C$_{60}$H$^{+}$ spectrum features weak bands near 9.25, 10.45 and 13.15 \textmu m (1090, 955 and 760 cm$^{-1}$), roughly characterized as ring breathing (9.25 \textmu m) and cage deformation (10.45 and 13.15 \textmu m). Finally, two stronger bands at 17.7 and 19.1 \textmu m (565 and 525 cm$^{-1}$) are close to the two main bands of neutral C$_{60}$ (17.4 and 18.9 \textmu m), with red-shifts relative to C$_{60}$ of 10 and 5 cm$^{-1}$, respectively. This is on the order of the bandwidth in the interstellar emission spectra.\cite{cami} 

\begin{figure*}[htp]
\begin{center}
\includegraphics[width=0.4\textwidth]{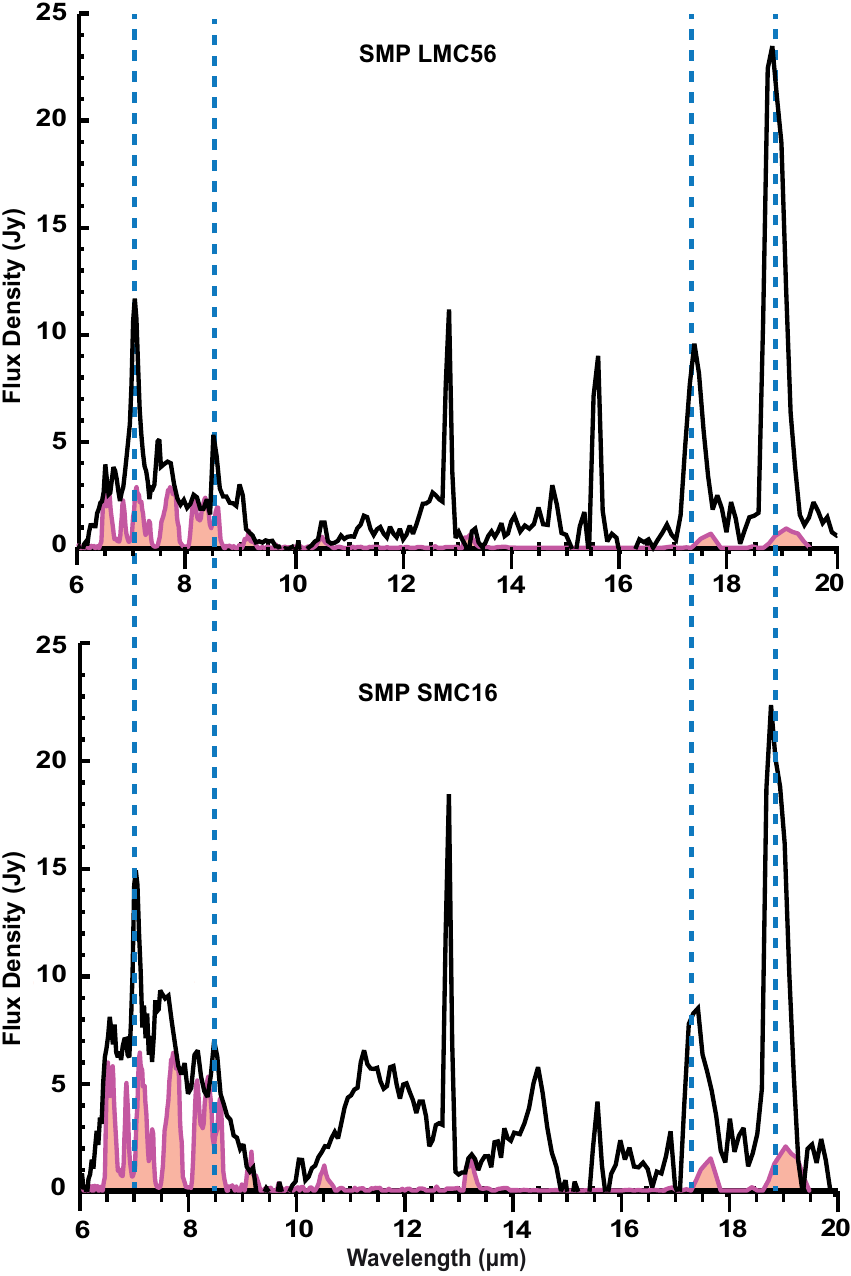}
\caption{\textbf{Comparison of the laboratory spectrum of C$_{60}$H$^{+}$ (pink) with the Spitzer IRS emission spectra  from the SMP SMC16 and LMC56 planetary nebulae.\cite{BernardSalas2012, Candian2019} Dashed lines indicate band positions for neutral C$_{60}$ at 7.0, 8.5, 17.3 and 18.9 $\mu$m. }}
\label{fig:Fig5}
\end{center}
\end{figure*}
    
C$_{60}$H$^{+}$ possesses a single CH-stretch mode with a computed integrated intensity of 27 km/mol, which we were unable to detect, probably due to the lower laser power available in this frequency range. As detailed in the Methods section, we estimate this band to occur at 2850 cm$^{-1}$, which deviates from the 100 cm$^{-1}$ broad CH stretch feature centered at 2910 cm$^{-1}$ reported for solid hydrogenated fullerenes.\cite{stoldt2001} On the other hand, this position coincides roughly with aliphatic (not aromatic) CH stretches  observed in gas-phase spectra of hydrogenated PAHs.\cite{maltseva2018} KBr pellet spectra of fulleranes (C$_{60}$H$_x$ with $x\approx 30$) show broad absorption features between 2800 and 2940 cm$^{-1}$.\cite{iglesiasG2012}

Since the first spectroscopic identification of C$_{60}$,\cite{cami} various astronomical objects have been particularly associated with high fullerene abundance.\cite{sellgren2010,c60GarciaH2010} Objects with low PAH emission are of particular interest to avoid confusion due to overlapping PAH bands. In Figure 4, the IR emission spectra of the planetary nebulae SMC16 and LMC56 \cite{BernardSalas2012,Candian2019} are overlaid onto our laboratory spectrum of C$_{60}$H$^+$ (see also Supplementary Figures 3 and 4). The two laboratory bands in the 17 -- 20 $\mu$m region fall within the bandwidth of the astronomical bands, although they are slightly shifted towards longer wavelengths with respect to the center position of the astronomical features, overlapping nearly exactly with the two bands in neutral C$_{60}$. The asymmetric lineshape of the astronomical features, with a shoulder towards longer wavelengths, may then be speculated to be due to the combined emission from C$_{60}$ and C$_{60}$H$^+$. Perhaps most striking is the close correspondence in the 6 -- 9 $\mu$m range: the laboratory spectrum falls closely within the envelope in the observational spectrum, with some of the individual peaks coinciding remarkably well with partially resolved structures on the envelope. In contrast to the two bands at longer wavelength, the entire 6 -- 9 $\mu$m emission feature cannot be explained by C$_{60}$ alone (dashed lines), nor does it appear to be due to PAHs (Supplementary Figure 2). The data presented here suggest that a mixture of protonated and neutral C$_{60}$ and higher fullerenes may form a plausible set of carriers for these emission features. A contribution from radical cation fullerenes\cite{fularaC60,berne2013,roithova2018} is also possible, but requires further investigation. Weaker bands in the remainder of the C$_{60}$H$^+$ spectrum are also consistent with the observational data. The 10.45 $\mu$m band appears visible in the LMC56 spectrum, and is consistent with the SMC16 spectrum, although overlapped by a broad emission plateau attributed to SiC.\cite{BernardSalas2012} Note that relative intensities in emission due to radiative cooling are amplified for bands at longer wavelengths with respect to bands at shorter wavelengths, while in photo-dissociation spectroscopy, this trend is opposite (see e.g.\ Figure 6 in Ref. \onlinecite{oomens2003}).    

The laboratory IR spectrum of C$_{60}$H$^+$ overlaid onto emission spectra of objects hypothesized to harbour high abundances of fullerenes appears to be supportive of Kroto's hypothesis that C$_{60}$H$^+$ is abundant. However, C$_{60}$H$^+$ alone cannot entirely explain the broad but partially resolved emission feature between 6 and 9 $\mu$m in these sources, but neither can available IR spectra of C$_{60}^{+\bullet}$.\cite{fularaC60,berne2013,roithova2018} We therefore speculate that a mixture of these species and their C$_{70}$ (and perhaps higher fullerene) analogues, is currently the most attractive explanation. New spectral data for C$_{70}$H$^+$ (currently ongoing in our lab) and C$_{70}^{+\bullet}$ are anticipated to further resolve this question. Higher spectral and/or spatial resolution astronomical data from future instruments such as the James Web Space Telescope (JWST), may resolve individual contributions more clearly.

\section*{Methods}
Experiments were performed in a modified 3-D quadrupole ion trap mass spectrometer (Bruker Amazon Speed ETD) coupled to the beamline of the Free-Electron Laser for Infrared eXperiments (FELIX). The modifications, including optical access to the trapped ions, have been described in detail in Ref.\ \onlinecite{martens2016}. 

To produce protonated C$_{60}$, an atmospheric pressure chemical ionization (APCI) source was employed, which is particularly efficient for less polar species, such as fullerenes \cite{McElvany1991,Anacleto1992} and polycyclic aromatic hydrocarbons \cite{Grosse2007,Lien2007}. Typically both protonated and radical cation species can be observed from APCI of these species.

C$_{60}$ (99+\,\%) was obtained from MER Corporation (USA). The powder is dissolved in toluene to a concentration of 1\,mmol, and this stock solution was further diluted in a 50/50\,\% methanol-toluene mixture to provide a final C$_{60}$ concentration of about 10\,\textmu M. This solution is infused at a flow rate of 8 \textmu l/min, nebulized with nitrogen at 3.5\,bar and 220\,$^{\circ}{\rm C}$, and vaporized at  450$\,^{\circ}{\rm C}$ . The fine droplets are sprayed towards the corona discharge needle. The potential difference between the end plate and capillary is 4500\,V and the corona current is set to 6000\,nA. Ions produced in the APCI source enter the vacuum of the mass spectrometer through a capillary, are guided by the ion-transfer optics and are then trapped in the radio-frequency quadrupole ion trap.

After mass isolation of the {\textit m/z} 721 ions in the trap, the ions are irradiated with tunable IR radiation from FELIX in order to measure the IR multiple-photon dissociation (IRMPD) spectrum of the trapped ions.\cite{oomens2003} The resonant multiple-photon absorption of IR photons by the ions leads to an increase of the internal energy and eventually to unimolecular dissociation.  Energy randomization via intramolecular vibrational redistribution (IVR) enables excitation of the molecules to the dissociation threshold, with the excitation energy being dissipated in all other vibrational degrees of freedom of the molecule. 

After irradiation of the trapped ion cloud, a mass spectrum is recorded and the normalized fragmentation yield ($Y$) is derived, taking into account the 35\% contribution of $^{13}$C$^{12}$C$_{59}^{+\cdot}$  to the signal at {\textit m/z} 721:
\begin{equation}
    Y = \frac{I_{m720}}{0.65 [I_{m720}+I_{m721}]}
\end{equation}
The fragment fluence $S$, which is proportional to the absorption cross section, is then obtained as
\begin{equation}
    S = -\ln (1-Y)
\end{equation}

\noindent
$S$ is then plotted as a function of the IR wavelength to provide a surrogate for the ion's IR spectrum. For every wavelength point, 5 mass spectra were averaged. $S$ was corrected linearly for wavelength-dependent variations in the laser pulse energy and for the irradiation time (number of applied laser pulses). From repeated scans of the spectrum, we estimate that dissociation yields are reproducible to within about 5\% . The FEL wavelength is calibrated with a grating spectrometer with an accuracy of $\pm$0.01\,\textmu m.

Spectra have been recorded in the vibrational fingerprint region between 6\,\textmu m and 25\,\textmu m. \cite{Oomens2006}. The FELIX pulses have an energy of up to 120 mJ and are produced at a 10 Hz repetition rate. Each pulse has a duration of about 7 \textmu s and consists of a series of micropulses spaced by 1 ns. The micropulses are Fourier-transform limited and have a bandwidth of 0.4\% of the IR wavelength.  

The high dissociation threshold of the protonated fullerene ions makes the application of photo-dissociation spectroscopy challenging, especially in the long wavelength range. In order to facilitate the on-resonance IR laser induced dissociation, the standard helium pressure settings of the ion trap, normally used to efficiently trap the ions, was reduced to its minimum value to minimize collisional deactivation of the ions during IR excitation \cite{martens2016}.  The degree of on-resonance dissociation is controlled by varying the laser pulse energy and by irradiating with a variable number of laser pulses (typically between 2 and 10) to prevent excessive depletion of the C$_{60}$H$^+$ precursor ion, which would manifest itself as saturation at strong absorption bands in the IR spectrum.

Computational investigations of the C$_{60}$H$^+$ system are carried out at the density functional theory (DFT) level using different combinations of functionals and basis sets. The hybrid B3LYP functional, known for its reliable performance in predicting IR spectra, was used in combination with a series of basis sets of increasing size: 3-21G, 4-31G (which has long been the standard in characterization of IR spectra for large PAHs\cite{bauschlicher2018,berne2013}) and 6-311+G(d,p). In addition, the non-hybrid BP86 functional (with the 4-31G basis set) was tested, analogous to strategies used in the computation of PAH IR spectra,\cite{bauschlicher2010} as well as the more recent dispersion-corrected functional M06-2X. 

All calculations employed the Gaussian16 software package as installed at the Cartesius supercomputer at SurfSARA, Amsterdam. Geometry optimizations were performed with the standard convergence criteria and vibrational spectra were computed within the harmonic oscillator approximation and  harmonic frequencies were scaled uniformly as recommended by Ref. \onlinecite{cccbdb}: 0.9679 for B3LYP/6-311+G(d,p), 0.9752 for B3LYP/4-31G, 0.965 for B3LYP/3-21G, 1.0033 for BP86/4-31G, 0.952 for M06-2X/6-311+G(d,p). The computational cost for the geometry optimization and the harmonic frequency calculation at the B3LYP/6-311+G(d,p) level of theory (2168 primitive gaussians) amounts to approximately 10 days on our 16-core computer cluster. For comparison with experiment, the computed stick spectra were convoluted with a Gaussian lineshape function with a full width at half maximum (FWHM) of 20 cm$^{-1}$, which is on the order of bandwidths typically observed in IRMPD spectra and which is believed to be due to a combination of laser bandwidth, rotational envelope and anharmonic broadening and shifting induced by the IRMPD process.\cite{oomens2003}

To estimate the frequency of the single CH-stretch band in C$_{60}$H$^+$, which we were unable to observe experimentally, we use the B3LYP/6-311+G(d,p) harmonic value and an appropriate frequency scaling factor. It is known that the scaling factor for CH stretch bands typically deviates from that for bands at longer wavelengths.\cite{bauschlicher2010} Therefore, we estimate this factor using B3LYP/6-311+G(d,p) computed spectra for pyrene and phenanthrene and comparing the CH-stretch frequencies with accurate gas-phase IR data.\cite{maltseva2016} A scale factor of 0.9610$\pm$0.0005 is  established, in agreement with Ref. \onlinecite{bauschlicher2010}. This gives a value of 2850 cm$^{-1}$ for the CH stretch mode of C$_{60}$H$^{+}$.

We note here that the astronomical spectra shown in this paper are emission spectra, resulting from radiative cooling of the species in the ISM after UV excitation and internal conversion.\cite{allamandola1989} Despite the fact that the IR emission process leaves its imprint on the appearance of the spectrum, as has been addressed in several modelling studies,\cite{schutte1993,saykally1998} spectral comparisons are often performed with linear vibrational spectra obtained through laboratory transmission spectra or quantum-chemical calculations. A generic redshift of 5 -- 10 cm$^{-1}$ or a frequency scaling is often applied to empirically correct for the small deviations. The laboratory spectrum of C$_{60}$H$^+$ reported in this contribution was obtained through IRMPD, which leaves its own imprint on the spectrum, which has been modelled in several studies as well.\cite{oomens2003,parneix2013} Again, deviations from linear vibrational spectra are small enough that spectral analyses are very commonly based on comparisons with linear spectra, as we do here in Figure 3.

The main challenge in correctly modelling IR emission and IRMPD spectra is that this requires parameters for line shifting and broadening as a consequence of anharmonicity, which are typically not available and are difficult, if not impossible, to determine accurately from quantum-chemical investigations. It is then interesting to note that although the processes of IR multiple-photon absorption and IR radiative cooling are different, they are dependent on the same molecular parameters, including these elusive mode-specific anharmonic parameters. To some extent, it may therefore be more appropriate to compare IR emission spectra with IRMPD spectra than with linear IR spectra. In any case, shifts are typically small, as is for instance evidenced by the good correspondence between our experimental IRMPD spectrum and the computed (linear) scaled harmonic spectrum of C$_{60}$H$^+$ at the B3LYP/6-311+G(d,p) level.

From these considerations, we suggest that the band positions in our experimental spectrum can be used directly in astronomical searches for C$_{60}$H$^{+}$. FWHM bandwidths of about 20 cm$^{-1}$ and slightly smaller at wavelengths near 20 $\mu$m are also expected to be close to those in an interstellar emission spectrum. Relative intensities may be somewhat less reliable as a consequence of non-linearities resulting from the IRMPD method; moreover a correction for the bias towards longer wavelengths in an emission spectrum needs to be taken into account.\cite{schutte1993,saykally1998,oomens2003}

\section*{Acknowledgements}
We gratefully acknowledge the expert support by the FELIX staff. This work is supported by the European MCSA ITN network "EUROPAH" (grant\# 722346) and the Dutch Astrochemistry Network (DAN-II, grant\# 648.000.030) of NWO. For the computational work, we acknowledge support by NWO under the "Rekentijd" program (grant\# 17603) and the SurfSARA staff. 

\section*{Author Contributions}
JP, JM and GB carried out the experiments, which were conceptualized by JO and GB. JP and JO wrote the manuscript with input from all other authors.  

\section*{Data availability}
Selected computational data are available in the Supplementary Information. The experimental data that support the findings of this study are available in the supplementary information.

\section*{Supplementary information} 
Computed frequencies and geometries, additional spectral comparisons, and data on exohedral versus endohedral protonation. Experimental IR spectral data of C$_{60}$H$^+$ as xy-file.

\section*{Competing interests} 
The authors declare no competing interests.

\end{document}